# Evidence of electronic phase separation in the strongly correlated semiconductor YbB$_{12}$

A. Azarevich[1], N. Bolotina[2,1], O. Khrykina[2], A. Bogach[1], E. Zhukova[3], B. Gorshunov[3],

A. Melentev[3], Z. Bedran[3], A. Alyabyeva[3], M. Belyanchikov[3], V. Voronov[1], N. Sluchanko[1]

[1]*Prokhorov General Physics Institute of the Russian Academy of Sciences, Vavilov str. 38,*

*Moscow 119991, Russia*

[2]*Shubnikov Institute of Crystallography, Federal Scientific Research Centre 'Crystallography and Photonics' of Russian Academy of Sciences, 59 Leninskiy prospect, Moscow, 119333, Russia*

[3]*Center for Photonics and 2D Materials, Moscow Institute of Physics and Technology (National Research University), 141700 Dolgoprudny, Moscow Region, Russia*

**Abstract.**

The studies of high-quality single-domain crystals of YbB$_{12}$ were carried out by the precise x-ray diffraction technique in combination with the low temperature polarized THz - infrared spectroscopy and accurate magnetotransport measurements. It has been shown for the first time that this archetypal strongly correlated system with a metal-insulator transition to a mysterious dielectric ground state with a metal Fermi surface (Science **362**, 65-69 (2018) and ibid **362**, 32-33 (2018)) is actually a *heterogeneous* compound in the regime of *electronic phase separation*. Changes in the configuration of the discovered dynamic charge stripes are investigated upon cooling, as a result, a conclusion is drawn in favor of a crossover between different patterns of the filamentary electronic structure penetrating the semiconducting matrix of YbB$_{12}$. We argue that the discovery of stripes in YbB$_{12}$ is fundamental, elucidating the nature of exotic dielectric state in Kondo insulators.

**I. Introduction.**

In the family of rare earth (RE) dodecaborides, YbB$_{12}$ is an archetypal strongly correlated electron system (SCES), which has been actively studied since the 1980s as a fluctuating-valence narrow-gap semiconductor [1-3]. Unstable valence 2.9–2.95 of Yb ions is accompanied with the metal-insulator transition (MIT), which is observed with decreasing temperature both in pure YbB$_{12}$ and in $R_{1-x}$Yb$_x$B$_{12}$ solid solutions with partial substitution of Tm or Lu for Yb (see [4] and references therein). Moreover, at low temperatures the insulator-to-metal transition is observed in YbB$_{12}$ under pressure [5] and in strong magnetic field [6, 7]. The mysterious insulating ground state in YbB$_{12}$ and similar SCES compound SmB$_6$ have attracted considerable interest from theorists, and various models have been proposed to explain the spectacular properties. Some of them are based on the coherent band picture where the energy gap forms due to the strong *d-f* hybridization [8, 9]. The inversion between 4*f* and 5*d* bands that accompanies this process was considered as an essential characteristic of the topological Kondo insulator [10, 11, 12]. Other models are based on a local picture, where the electrons contributing to the Kondo screening are captured by the local magnetic moments resulting in an excitonic local Kondo bound state [13]. Among the exotic electronic properties recently discovered in the archetypal SCES narrow-gap semiconductors SmB$_6$ and YbB$_{12}$ with a very unusual insulating ground state and metallic Fermi surfaces, we highlight (a) quantum oscillations of magnetization and resistivity, which are the fingerprints of normal metal, and (b) extraordinary gapless charge-neutral fermionic excitations [6-7, 14-16].

The above-mentioned models are based on the postulate of *stable B$_{12}$ nanoclusters* as the main structural elements of the RE dodecaborides $R$B$_{12}$, whose structure is similar to the *fcc* structure of NaCl when Na is replaced by $R$ and Cl is replaced by B$_{12}$ cuboctahedron (Fig. 1a). On the contrary, it was shown recently (see [17, 4] and references therein), that the cooperative Jahn-Teller (JT) instability of the boron framework (ferrodistortive effect) is among the main factors responsible for (*i*) the static and dynamic *fcc* lattice JT distortions and (*ii*) formation of dynamic charge stripes in the matrix of $R$B$_{12}$. These fluctuating charges (stripes) lead to nanoscale phase separation [18], which creates a *strong inhomogeneity in the electron density* (ED) distribution in the conduction band, combined with symmetry breaking in these highly entangled SCES [17, 4]. In the Yb-based RB$_{12}$ the JT instability of the boron sublattice is accompanied with the valence instability of Yb, so that fast on-site 4*f*-5*d* charge and spin fluctuations of ED can modify the dynamic charge stripe patterns in the model SCES [4]. The dynamic charge stripes (ED fluctuations with the frequency $\nu_S$ ~ 240 GHz [19]) were discovered previously in the Tm$_{0.19}$Yb$_{0.81}$B$_{12}$ even at room temperature [4, 19], however, the possibility of existence of these singularities in the parent narrow-gap semiconductor YbB$_{12}$ was not investigated up to now.

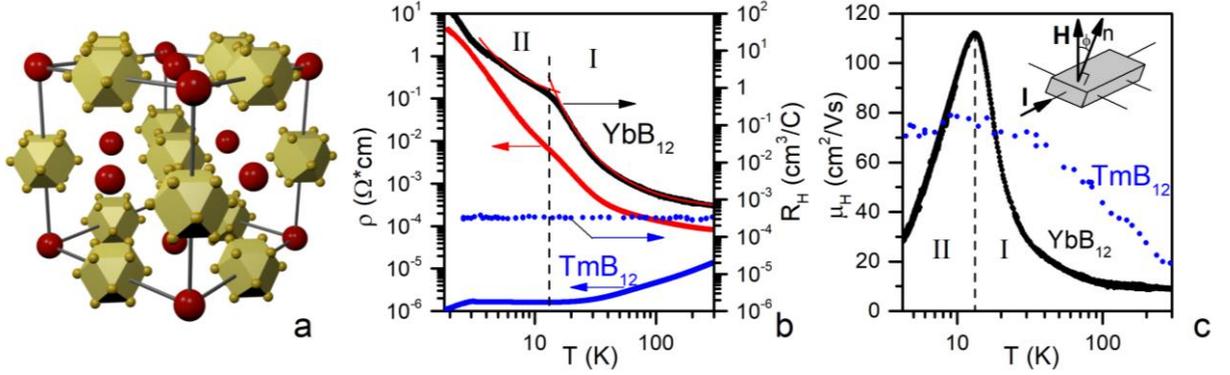

**Fig. 1.** (a) Crystal structure of $R$B$_{12}$, red and khaki balls mark the RE and boron atoms, respectively. Temperature dependences of (b) resistivity $\rho(T)$ and Hall coefficient $R_H(T, H\sim 5$ kOe) and (c) Hall mobility $\mu_H(T)= R_H(T)/\rho(T)$ of charge carriers in YbB$_{12}$ and TmB$_{12}$ (for comparison). Solid lines in (b) demonstrate the activation behavior of the Hall coefficient. Inset in (c) explains the idea of measuring the angular $\phi$-dependence of the magnetoresistance; vector **n** is perpendicular to the sample surface. Roman numerals I and II denote the charge transport regimes (see text).

To shed more light on the nature of the ground state, the driving mechanisms of fluctuating charges and their spatial patterns in the Yb-based mixed-valence SCES, we carried out (*i*) x-ray diffraction (XRD) studies of high-quality YbB$_{12}$ single crystals, (*ii*) measurements of the low-temperature polarized terahertz response, and (*iii*) accurate charge transport experiments. Clear evidence of charge stripes was demonstrated by the x-ray structure analysis at $T\sim 100$K, both for YbB$_{12}$ and TmB$_{12}$. We show that it is the discovered filamentary structure of the ED that is at the origin of an excitation detected at low temperatures in the terahertz spectra of AC conductivity and permittivity for polarization of the electric field vector **E** of the probing radiation oriented along to the stripes, E⊥[110]; no such excitation was observed in case of **E**∥[110]. Additional evidence is presented by the anisotropy of the magnetoresistance, measured for the magnetic field oriented along or across the dynamic stripes. Emergence of stripes in YbB$_{12}$ is a fundamental finding, indicating exotic dielectric state in the Kondo insulator with metallic Fermi surface.

**II. Experimental details.**

X-ray data on high quality YbB$_{12}$ single crystal (the same as in [20]) were collected on an Xcalibur diffractometer (Rigaku Oxford Diffraction) equipped with an EOS S2 CCD detector, using Mo$K_\alpha$ radiation, λ = 0.71073 Å. The sample was cooled using a Cobra Plus cryosystem (Oxford Cryosystems) with an open flow of cold nitrogen directed at the sample (for details, see [20]). A rectangular single-crystalline sample 4×0.2×0.2 mm$^3$ in size was prepared from the same batch and used to study the resistivity and Hall resistance in a five-terminal scheme with a direct current (DC) commutation at temperatures 4.2–300 K in a magnetic field up to 80 kOe. As shown in the inset to Fig. 1c, the long axis [0-11] of the sample coincides with the direction of current ***I***, oriented across the external magnetic field ***H***, and ϕ is the angle between ***H*** and [100] axis of the crystal. The measurements of the magnetoresistance were carried out by rotating the crystal around the [01-1] axis with a step of Δϕ = 1.8°. Broad-band terahertz-infrared (frequencies ν ≈ 10-1000 cm$^{-1}$) spectra of conductivity and dielectric permittivity were obtained by using terahertz time-domain and infrared Fourier-transform spectrometers as discussed in detail in Supplementary Information section [21].

**III. Results and discussion.**

**III.1. Temperature dependences of resistivity and Hall effect.** The resistivity ρ(*T*) and Hall coefficient $R_H(T)$ of YbB$_{12}$ measured in the range 4.2–300 K are shown in Fig. 1b demonstrating the metal-insulator transition during cooling. For comparison, Fig. 1b also presents the charge transport characteristics of metallicTmB$_{12}$. Both the resistivity and Hall coefficient of YbB$_{12}$ increase drastically with decreasing temperature. In the two temperature intervals, I (20-300 K) and II (5-12 K), an activation type behavior is evidenced from the analysis of these two charge transport characteristics (see, for example, solid lines in Fig.1b), assuming two energy scales that correspond to the indirect gap $E_g/k_B ≈ 216$ K and to the intra-gap excitation $E_S/k_B ≈15$ K ($k_B$ is the Boltzmann constant) estimated from the Hall effect data in accord with previous results [4,1]. Detailed analysis of Hall mobility $\mu_H(T)= R_H(T)/\rho(T)$ will be presented elsewhere. The two regimes I and II are separated in YbB$_{12}$ by the strong peak at $T_S$~15 K on the $\mu_H(T)$ curve (Fig.1c). Note also that the crossover to the low temperature coherent regime of charge transport at $T_S$ was attributed in [4] to emergence of large size clusters of the manybody states in the Yb-based narrow gap semiconductors. Moreover, it was argued [4], that the filamentary structure of the conduction electrons is composed of the charge stripes in the RE dodecaborides, and $T_S$ is the measure of the electron density fluctuations: $k_B T_S \sim h\nu_S \sim 1$ meV (*h* is the Planck constant).

**III.2. Difference electron density distribution.** Crystal structures of Kondo insulator $YbB_{12}$ [20] and metal $TmB_{12}$ [17] were studied earlier in the range 85–293 K based on x-ray diffraction data. Weak static Jahn–Teller distortions of the cubic lattice were detected at all temperatures [17], but this did not require a revision of the highly symmetrical structural model. Both structures were refined at all temperatures in the cubic space group $Fm\bar{3}m$. The intensities of diffraction reflections measured at a temperature of 107 K, and the results of refinement of the structural models of $TmB_{12}$ and $YbB_{12}$ are used in this work to calculate difference Fourier syntheses of the electron density $\Delta g$ [21]. As seen in Fig. 2, positive stripe-like electron density residues are observed along the [110] direction in both $YbB_{12}$ and $TmB_{12}$. The stripes, which are more pronounced in $YbB_{12}$, are observed not only in rows with metal atoms, but also in rows consisting only of $B_{12}$ cuboctahedrons. In the case of $TmB_{12}$ the ED distribution in stripes is wider and located mainly near the boron atoms. The axes in the crystal indicated in Fig. 2a correspond to the directions in Fig. 4 below.

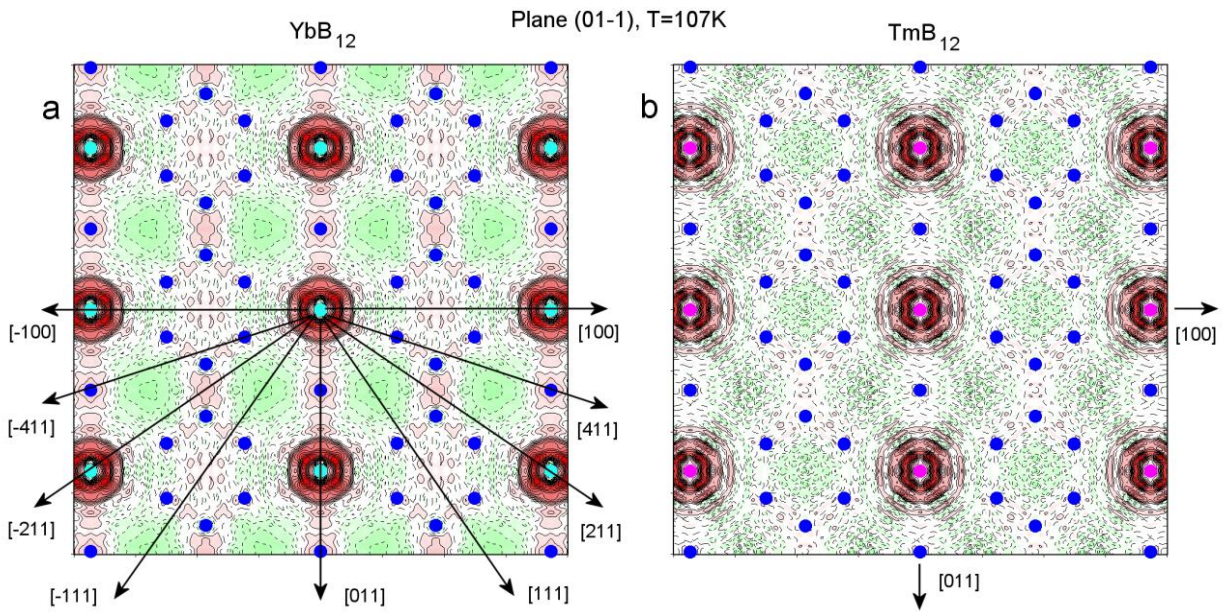

**Fig.2**. Difference Fourier synthesis of electron density $\Delta g$ in $YbB_{12}$ (a) and $TmB_{12}$ (b) crystals studied at temperature 107 K. The distribution $\Delta g$ in a layer 1 Å thick was obtained by summing over parallel (01-1) sections with a step of 0.05 Å normalized to the layer thickness. The blue circles indicate the boron positions in the layer or near the layer; the light blue and pink circles mark the Yb and Tm positions, respectively. Shades of green and red indicate areas on the map with negative and positive values of $\Delta g$, respectively. Contour intervals are 0.2 e/Å$^3$.

**III.3. Dynamic conductivity.** Fig. 3 shows the terahertz-infrared spectra of conductivity and real and imaginary permittivity of the $YbB_{12}$ single crystal measured at different temperatures in two polarizations with AC electric field oriented in the sample surface plane (1-11) along and across the [110] axis . The spectra look similar to those measured previously with unpolarized radiation [22]. The conductivity spectra for both polarizations practically coincide at intermediate temperatures (see, for example, the spectrum measured at 80 K in Fig.3a). Note, that at 80 K the $\sigma(\nu)$ behavior is typical for conductor: the Drude-type conductivity increases below 100 cm$^{-1}$ toward the DC value.  Upon cooling, strong drop in the conductivity spectrum is observed below 300 cm$^{-1}$ (≈40 meV) evidencing opening of a direct gap (indicated by vertical arrows in Fig.3a,c) in the electron density of states (see also [22, 23]). At $T$=10 K two more features are distinguished at frequencies below 300 cm$^{-1}$, as seen in the three panels of fig. 3: (a) a shoulder at about 150 cm$^{-1}$ in $\sigma(\nu)$ and $\varepsilon''(\nu)$ spectra and (b) a peak at 60-70 cm$^{-1}$ for **E**⊥[110]; note that no similar peak is seen for the case of **E** ∥ [110].

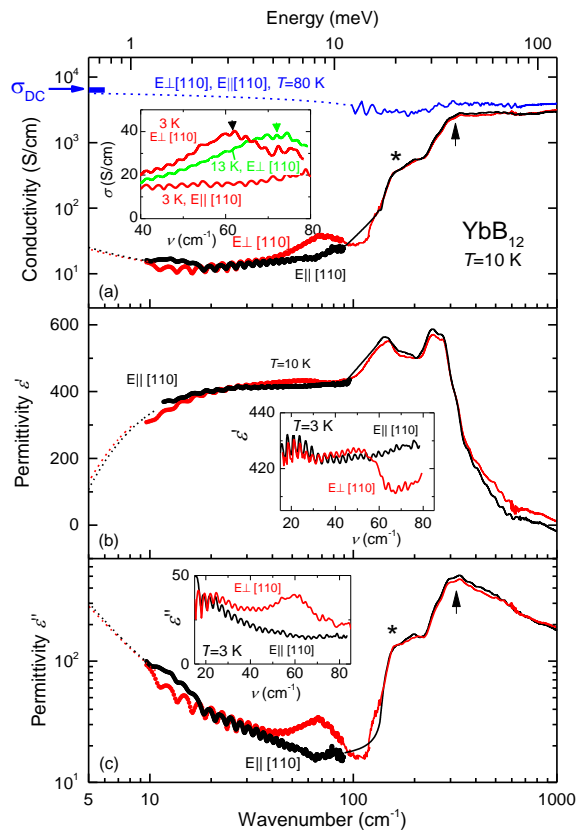

**Fig.3.** Terahertz (dots) and infrared (lines) spectra of conductivity (a) and of real (b) and imaginary (c) dielectric permittivity of YbB$_{12}$ single crystal measured at different temperatures and polarizations as indicated. Dashed lines show results of Kramers-Kronig analysis of reflectivity spectra. Arrows and stars in panels (a), (c) indicate direct gap and collective mode in the spectra, respectively. Insets show terahertz spectra measured at $T$=3 K for two different polarizations. Horizontal bar in panel (a) indicated DC conductivity measured at $T$=80 K. Small oscillations at terahertz frequencies (below 100 cm$^{-1}$) are caused by standing waves due multiple reflections of the radiation within the cryostat.

From the inset of Fig.3a one can see that the position of the peak shifts from ≈ 72 cm$^{-1}$ at $T$=13 K to ≈ 62 cm$^{-1}$ at $T$=3 K. We associate the origin of the peak with plasmonic oscillations of charge carriers that are delocalized within the stripes elongated transverse to the [110] direction and driven by AC electric field of the terahertz radiation. Similar plasmonic excitations had been previously observed in experiments on carbon nanotubes (see [24], for example). Within such assumption and following the lines used in [24], we estimate the distance between which the charge carriers are localized, i.e., stripes length, as $L=V_p(\pi v_p)^{-1}$, where $v_p$ is the position of the plasmon peak and $V_p$ is the plasmon velocity that is several times larger than the Fermi velocity, $V_p \approx 4V_F$ [25]. Taking the stripes energy of ≈1 meV we estimate the Fermi velocity of the carriers as $V_F \approx 1.9 \times 10^6$ cm/s. With $v_p$=72 cm$^{-1}$ (at 13 K) and $v_p$=62 cm$^{-1}$ (at 3 K) the average stripes length is evaluated increasing from≈100 Å at $T$=13 K to ≈116 Å at $T$=3 K. In [22], a clear maximum in the AC conductivity spectra of YbB$_{12}$ was also observed in an unpolarized radiation, but at another frequency of 20 cm$^{-1}$. Lower frequency can indicate larger stripes length in the crystal measured in [22].

**III.4. Magnetoresistance anisotropy.** As established earlier [17, 4], considering the distribution of the ED estimated from XRD data in YbB$_{12}$ allows to explain the anomalies in the angular dependences of the low-temperature magnetoresistance (MR) by the scattering of charge carriers on the dynamic charge stripes both in the nonmagnetic reference dodecaboride LuB$_{12}$, and in the paramagnetic state of $R$B$_{12}$ with trivalent magnetic RE ions ($R$ = Ho, Er and Tm). In more detail, the inhomogeneous distribution of ED (nanoscale phase separation) in $R$B$_{12}$ ($R$ = Lu, Ho, Er and Tm), induced by the cooperative JT instability of B$_{12}$ clusters, leads to an additional positive component in both the MR and the Hall resistivity due to the interaction of the filamentary structure of fluctuating charges with an external magnetic field. According to [4], this interaction has a maximum if the magnetic field is transverse to the dynamic stripes, which are directed along the [110] axis in all dodecaborides with R$^{3+}$ ions. Magnetic field and angular dependences of MR $\Delta\rho/\rho = (\rho(H)-\rho(0))/\rho(0)$ at $T$=4.2 K are presented in Fig. 4 for TmB$_{12}$ and YbB$_{12}$. As shown in Fig. 4b, for TmB$_{12}$ the MR reaches its maximum for **H** ∥ [100], but in YbB$_{12}$ the low-temperature MR anisotropy changes its polarity (Fig. 4d). The sharp change in the MR anisotropy can be explained by the reorientation below $T_S$ ~ 15 K of the ED stripes in YbB$_{12}$ from [110] to [100], which is caused by strong 4$f$–5$d$ charge and spin fluctuations at the Yb-ions. As a result, we can assume the appearance of new configurations of dynamic charge stripes in YbB$_{12}$ additionally to those caused by the cooperative JT instability of the rigid boron frame.

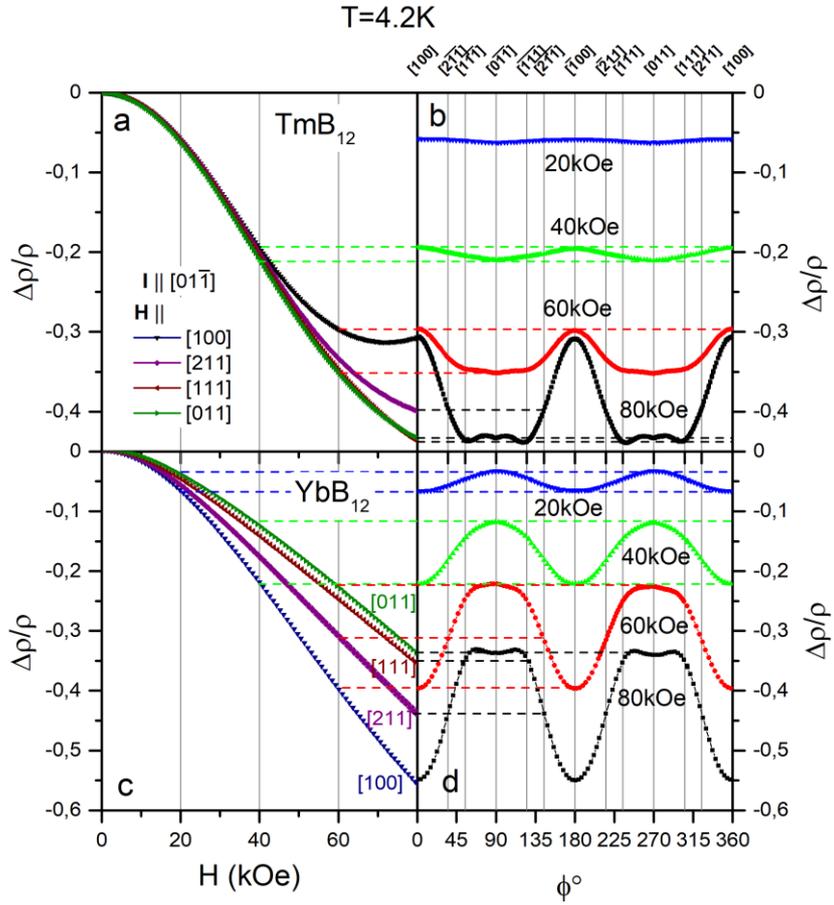

**Fig.4.** Magnetic field (a), (c) and angular (b), (d) dependences of MR $\Delta\rho/\rho=(\rho(H)-\rho(0))/\rho(0)$ at $T = 4.2$ K for $TmB_{12}$ and $YbB_{12}$, correspondingly. Each angle $\phi$ corresponds to the crystallographic axis directed along **H** when the crystal rotated in the (1-10) plane through the angle $\phi$ around the [01-1] axis (see the schematic view in the inset in Fig. 1c).

**III.5. Changes in the stripe patterns in $YbB_{12}$.** Note that the XRD results in Fig. 2 indicate the [110] direction of the dynamic charge stripes in both $YbB_{12}$ and $TmB_{12}$ above 100 K, at least. Below $T_S \sim 15$ K (to the left of the mobility maximum in Fig. 1c), new patterns of the ED filamentary structure dominate, which are determined mainly by the $4f$–$5d$ charge fluctuations at the Yb positions, and, consequently, the MR anisotropy changes polarity (Fig. 4b, 4d). Thereby, low-temperature charge fluctuations become transverse to [110], which leads to the appearance of polarization effects in the terahertz conductivity of $YbB_{12}$, where a plasmon peak is observed at 60–70 cm$^{-1}$ for polarization **E**⊥[110] (Fig. 3). The transformation of $YbB_{12}$ stripe patterns in the temperature range 15–100 K correlates very well with dramatic changes above $T_S \sim 15$ K in the spectra of magnetic excitations detected in [26] by the inelastic neutron scattering (INS). Indeed, it was found [26] that, with increasing temperature, the magnetic excitations begin to transform: the spin gap is gradually filled with a broad quasi-elastic signal

(half-width $\varGamma_{sf}/2 \sim 10$ meV), while three narrow INS peaks below 40 meV are concurrently suppressed. The transformation to the high-temperature regime of spin fluctuations ends near $T_{sf} \sim \varGamma_{sf}/2 \sim 100$ K, and the energy of spin (and charge) fluctuations is considered as the boundary between these two regimes in YbB$_{12}$ with an intermediate valence of Yb-ions. Similarly to the ED patterns at $T \sim 100$ K (Fig.2), the low-dimensional character of the spin fluctuation spectrum was revealed for YbB$_{12}$ at liquid helium temperatures [26]. Indeed, the inelastic peak in the magnetic excitation spectrum of YbB$_{12}$ at ~15 meV was attributed to antiferromagnetic correlations with the wave vector $\boldsymbol{q}= (1/2, 1/2, 1/2)$, and the correlation lengths were estimated both perpendicular and parallel to [110] direction providing the values $\xi_\parallel = (5.4 \pm 1.4)$ Å within the (001) planes and $\xi_\perp = (3.4 \pm 1.1)$ Å between the (001) planes [26]. The INS results confirm the low-dimensional anisotropy in this strongly correlated electron system with *fcc* crystal structure, and the value $\xi_\parallel \sim 5.4$ Å correlates very well with the shortest Yb-Yb distance in the *fcc* lattice.

It is worth noting also, that a strong increase of the signal of electron spin resonance (ESR) was observed previously in YbB$_{12}$ [27], and the temperature dependence of the ESR amplitude was found to be close to exponential increase with a characteristic temperature ~18 K, which is close to $T_S \sim 15$ K. The authors [27] conclude that the ESR results can be understood assuming the existence of Yb$^{3+}$ ion pairs coupled by isotropic exchange interaction, which also interact with neighboring Yb-pairs connecting these dimers in the nanosize channels.

**IV. Conclusion.**

The dynamic charge stripes have been discovered in the archetypal strongly correlated electron systemYbB$_{12}$ with metal-insulator transition by precise XRD data analysis, polarized terahertz spectroscopy and the charge transport measurements. At T ~ 100 K, near the temperature of spin fluctuations $T_{sf}$, YbB$_{12}$ exhibits stripe patterns oriented along the [110] direction, with the corresponding filamentary electron density structure found not only between rare-earth metal atoms, but also in rows consisting of B$_{12}$ cuboctahedrons. On cooling well below $T_{sf}$ at temperatures close to $T_S \sim 15$ K a crossover to the coherent regime of the charge transport takes place. It is accompanied by setting up of the low dimensional (1D or 2D) spin and charge fluctuations and formation of large size (>100 Å) clusters in the nanoscale filamentary structure of conduction electrons oriented transverse the [110] direction within the semiconducting matrix of YbB$_{12}$. We propose the formation of stripes in YbB$_{12}$ as a fundamental phenomenon, elucidating genesis of exotic dielectric state in the Kondo insulator with metallic Fermi surface.

**Acknowledgements**. The authors are grateful to A.P. Dudka for collecting and processing data using ASTRA, S.V. Demishev, V.V. Glushkov, N. Yu. Shitsevalova and V. B. Filipov for helpful discussions, K.M. Krasikov for experimental assistance and Z. Gagkaeva for assistance with spectral analysis. This work was partially supported by the Ministry of Science and Higher Education within the State assignment FSRC 'Crystallography and Photonics' RAS in part of structural analysis. The study of properties of the $YbB_{12}$ and $TmB_{12}$ crystals was carried out with the support of the Russian Science Foundation (Project No. 22-22-00243).